\documentclass[journal]{IEEEtran}

\usepackage{cite}
\usepackage{xcolor}

\ifCLASSINFOpdf
  \usepackage[pdftex]{graphicx}

   \graphicspath{{../pdf/}{../jpeg/}}

   \DeclareGraphicsExtensions{.pdf,.jpeg,.png}
\else

   \usepackage[dvips]{graphicx}

   \graphicspath{{../eps/}}

   \DeclareGraphicsExtensions{.eps}
\fi

\usepackage{amsmath}
\usepackage{amssymb}

\usepackage{subcaption}

\usepackage{float}

\usepackage{tikz}
\usepackage{textcomp}
\usepackage{hyperref}

\newcommand\copyrighttext{%
  \footnotesize 
This is the author's version of an article that has been published in IEEE Communications Letters, 24(12), 

pp. 2974 - 2978, 2020. Changes were made to this version by the publisher prior to publication.

The final version of record is available at \href{http://dx.doi.org/10.1109/LCOMM.2020.3019515}{http://dx.doi.org/10.1109/LCOMM.2020.3019515}.

Copyright (c) 2020 IEEE. Personal use of this material is permitted. For any other purposes, permission must 

be obtained from the IEEE by emailing pubs-permissions@ieee.org.}
\newcommand\copyrightnotice{%
\begin{tikzpicture}[remember picture,overlay]
\node[anchor=north,xshift=50pt,yshift=-7pt] at (current page.north) {\fbox{\parbox{\dimexpr0.7\textwidth-\fboxsep-\fboxrule\relax}{\copyrighttext}}};
\end{tikzpicture}%
}

\begin{document}

\title{GoT: decreasing DCC queuing for CAM messages}

\author{Oscar~Amador, Ignacio~Soto, Manuel~Urueña,
        and~Maria~Calderon
\thanks{O. Amador, I. Soto, and M. Calderon are with the Departmento de Ingeniería Telemática, Universidad Carlos III de Madrid, 28911 Leganés, Spain (e-mail: oamador@pa.uc3m.es; isoto@it.uc3m.es; maria@it.uc3m.es).}
\thanks{M. Urueña is with the Escuela Superior de Ingenieros y Tecnología, Universidad Internacional de la Rioja, 26006 Logroño, Spain (email: manuel.uruena@unir.net).}
\thanks{This work was partially supported by the Spanish Ministerio de Economía y Competitividad through the Texeo project (TEC2016-80339-R).}}

\markboth{IEEE Communications Letters}%
{Amador \MakeLowercase{\textit{et al.}}: GoT: decreasing DCC queuing for CAM messages}


\maketitle
\copyrightnotice
\begin{abstract}
Vehicular networks use Decentralized Congestion Control (DCC) mechanisms to operate effectively, but this mechanism may introduce queuing delays. Freshness of Cooperative Awareness Messages (CAMs) is critical for their usefulness. In this letter we explore how the presence of other types of traffic additional to CAMs, even with lower priorities, has an impact on the freshness of CAM messages due to DCC queuing. Finally, we propose Generate-on-Time (GoT), which is a simple mechanism that reduces DCC queuing delays for CAM messages without introducing any downside in other performance metrics.
\end{abstract}


\IEEEpeerreviewmaketitle

\section{Introduction}
Cooperative awareness (CA) services are one of the cornerstones of Intelligent Transport Systems (ITS). It is through these services that ITS stations learn about the position, heading, speed, and contextual information about their neighbors on the road. The European Telecommunication Standards Institute has defined a standard for a Cooperative Awareness basic service (ETSI EN 302 637-2) \cite{etsiCAM}, which specifies Cooperative Awareness Messages (CAM). A CAM contains information regarding the status (e.g., time, position, activated systems) and attributes (e.g., dimensions, vehicle type and role) of a generating station. CAM messages are generated periodically, and their frequency is affected by changes in the status of the generating station as well as by channel occupation.

\begin{figure}[t]
	\centering
	\includegraphics[width=190px,clip,keepaspectratio]{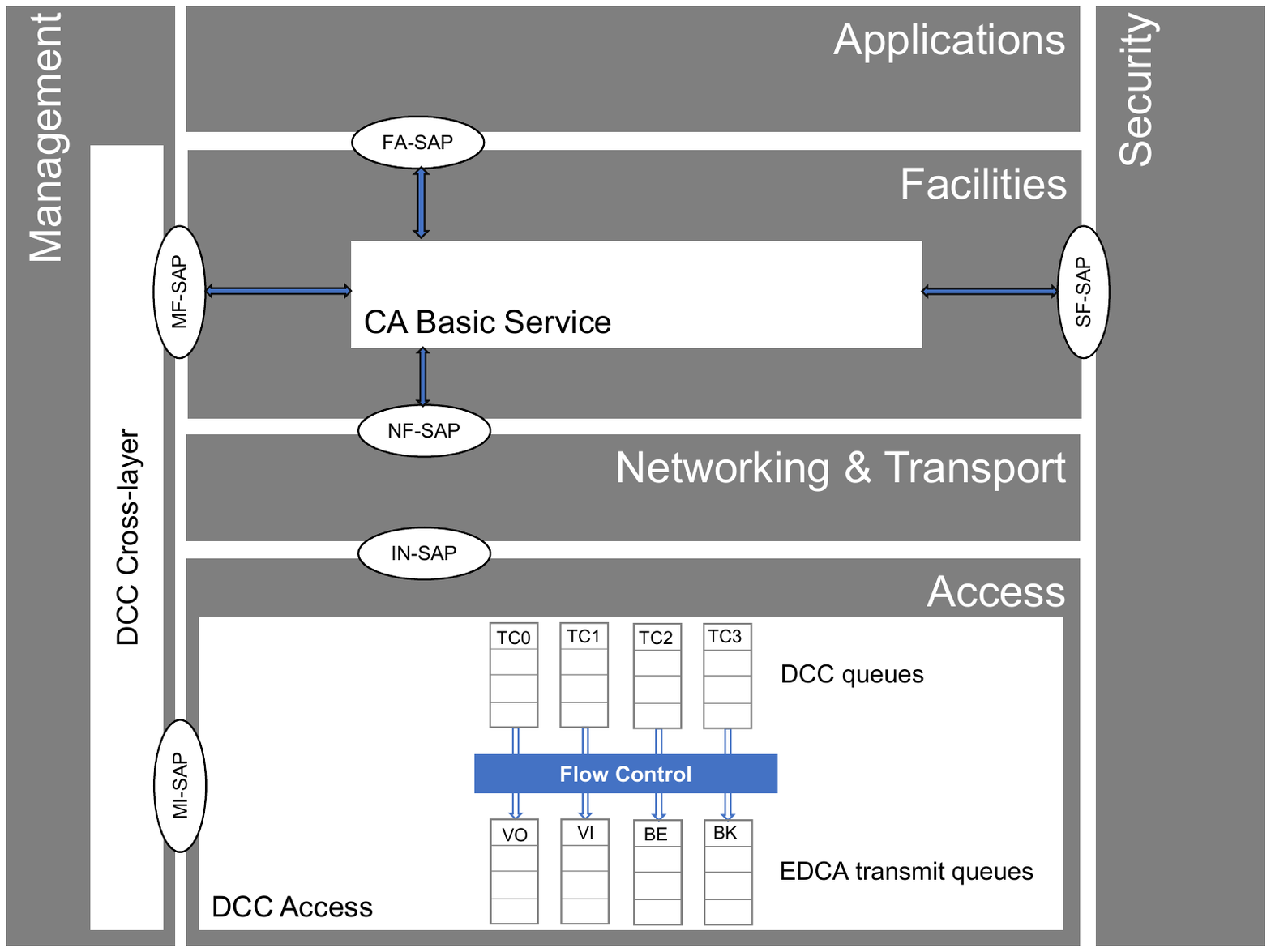}
	\caption{ETSI ITS Architecture: DCC cross-layer and the CA basic service.}
	\label{fig:camdccarch}
\end{figure}

Since the objective of a CA service is to keep neighbors aware of a vehicle's status, the \textit{validity} and \textit{recentness} of the information in CAM messages is paramount. There are factors that affect these two attributes, such as collisions ---that lead to messages being received at longer intervals---, and end-to-end delay, which is the time difference between the generation of a CAM in a station and its consumption in the Cooperative Awareness service of a remote neighbor.
Extensive work has been conducted on evaluating the performance of CA services. The bulk of the work has been focused on metrics pertaining losses and collisions, such as \cite{CAMboban}, which evaluates metrics such as packet delivery and neighborhood awareness ratios; while \cite{CAMerlangen}, \cite{CAMe2e} also measured latency (i.e., end-to-end delay) for CAM messages in urban and highway scenarios\textcolor{black}{. Nonetheless, these two publications} only consider CAM traffic, with messages being generated at a fixed rate of 10 Hz. However, CAM generations (as defined by ETSI) occur at dynamic rates, and they are influenced by Decentralized Congestion Control (DCC) mechanisms. The ETSI ITS-G5 protocol stack considers the use of DCC at different layers \cite{etsiNewDcc}, one of which is Facilities, where the Cooperative Awareness service is located, and the rate at which CAM message generation occurs is controlled by a cross-layer DCC \textcolor{black}{mechanism~\cite{etsiCross}, as shown in Fig.~\ref{fig:camdccarch}.}

\textcolor{black}{In the past, several publications~\cite{Sepulcre2016}~\cite{dccHarri} have studied} the problem of coordination between the different layers at which the DCC mechanism operates. The work in~\cite{Sepulcre2016} proposes an integrated congestion control solution that takes into account the requirements of CA applications and the experienced channel load. The work in~\cite{dccHarri} proposes that DCC functionalities should be performed only at the Facilities layer. However, these previous works do not propose how to improve the coordination between Access and Facilities layers in the ETSI architecture as our paper does. On the other hand, several past works have analyzed the performance of congestion control protocols considering the coexistence of distinct V2X ITS services~\cite{Harri2018,gunther2018,dccHarri}. Several of these works have considered the coexistence of messages with different priorities (i.e., CAM and lower-priority traffic) \cite{Harri2018}~\cite{gunther2018}; however, they do not analyze how this low priority traffic can impact the freshness of CAMs as our work does.

\textcolor{black}{Our previous work \cite{Letter} \cite{access} had the aim of evaluating the Adaptive Approach of }the ETSI Decentralized Congestion Control mechanism and proposed a variation that improves the performance in rapidly-changing scenarios. In \cite{access}, we found an effect of multi-traffic on CAM latency, using scenarios with realistic CAM generation (i.e., based on vehicle dynamics and the rate allowed by DCC) and another type of traffic with lower priority coming from the same station. Since the transmission of different types of traffic is controlled by a single DCC mechanism, following the rules from \cite{etsiCAM}, the transmission of another message might delay the transmission of a CAM that is generated afterwards, making it wait at the DCC queue. Furthermore, this desynchronizes the DCC mechanism at the Access layer and the CA service, leading to further queuing delays. 

The main contribution of the present work is a mechanism --- Generate-on-Time (GoT) --- to reduce waiting times of CAM messages at DCC queues and, thus, end-to-end delay. GoT delays the generation of CAMs until they can be sent, without altering the rate at which they are transmitted, by synchronizing the cross-layer DCC mechanism. The work is divided as follows: in section \ref{sec:e2e}, \textcolor{black}{an analysis} of CAM end-to-end delay in multi-traffic scenarios is presented, considering the current ETSI CAM standard \cite{etsiCAM} and GoT, our proposed solution. Section \ref{sec:simulation} presents results of simulations on fixed and dynamic scenarios to evaluate the performance of ETSI CAM and GoT. Finally, conclusions are presented in section~\ref{sec:conclusion}.

\section{End-to-end delay in CAMs}
\label{sec:e2e}
\subsection{Overview of CAM generation rules}

\textcolor{black}{Following the rules established by \cite{etsiCAM}, the CAM generation interval is limited within a range from 0.1\,s to 1\,s by four parameters: 
\begin{itemize}
	\item $T\_Elapsed$ (i.e., time since last CAM generation),
	\item vehicle dynamics (i.e., shifts in position, acceleration, and heading),
	\item $T\_GenCam\_DCC$ (i.e., lower limit of the CAM generation given by the DCC mechanism), and
	\item $T\_GenCam$ (i.e., upper limit for the generation interval).
\end{itemize} 
A CAM generation is triggered after $T\_Elapsed$ greater than or equal to $T\_GenCam\_DCC$ by either of two conditions: 
\begin{enumerate}
	\item if vehicle dynamics have exceeded certain thresholds, or
	\item if $T\_Elapsed > T\_GenCam$.
\end{enumerate}
Where $T\_GenCam$ is $T\_Elapsed$ for the last CAM generated by condition 1. After three CAMs have been generated by condition 2, $T\_GenCam$ is set to 1\,s}.

When an outgoing CAM message reaches the Access layer, it is placed at the DCC queue corresponding to \textcolor{black}{Traffic Class 2 (TC2)~\cite{etsiDP}}. There, the congestion control mechanism dequeues messages according to the priority of their \textcolor{black}{Traffic Class (i.e., TC0 to TC3, with TC0 having the highest priority)}. When the DCC mechanism indicates that it is time for the next transmission (i.e., $t_{go}$ from \cite{etsiNewDcc}), the message is sent to the lower-layer transmit queues. In the presence of a single type of traffic, $t_{go}$ and $T\_GenCam\_DCC$ will usually coincide, which means that a generated CAM will not have to wait in the DCC queues and it will only be delayed by the medium access control mechanism (i.e., EDCA) until it is finally transmitted.

\subsection{Analysis of CAM end-to-end delay with multi-traffic}

\begin{figure}[t]
	\centering
	\includegraphics[width=200px,clip,keepaspectratio]{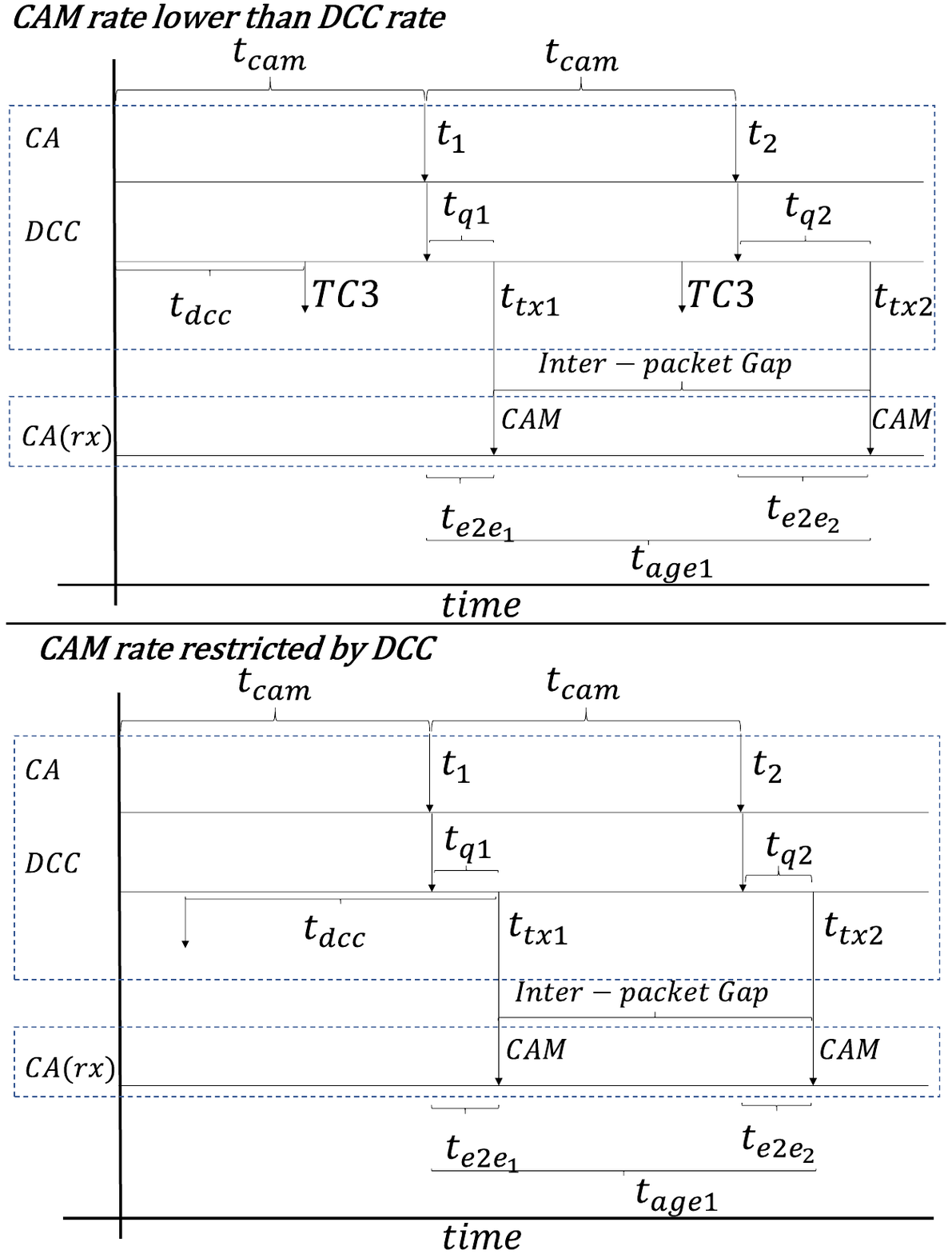}
	\caption{CAM generation under rules defined in \cite{etsiCAM}}
	\label{fig:dia1}
\end{figure}

Fig. \ref{fig:dia1} illustrates how CAM messages are generated, enqueued, and transmitted in a vehicle and received in a remote station when CAM coexists with lower-priority traffic \textcolor{black}{(e.g., multihop DENM messages, and other data traffic~\cite{etsiGeoNetworking})}. The interval at which CAMs are generated is expressed as $t_{cam}$, and $t_{dcc}$ represents the interval at which DCC allows a vehicle to send traffic. The upper part of the diagram shows the case when CAM messages are generated at rates lower than those allowed by the cross-layer DCC mechanism, and the lower part represents what happens when CAM message rate is restricted by the DCC mechanism (i.e., dynamics trigger a CAM every time DCC allows a generation). 

As shown in Fig.~\ref{fig:dia1}, CAM messages are generated every $t_{cam}$. At $t_{1}$, a CAM is generated (with a timestamp and information corresponding to $t_{1}$), but since a \textcolor{black}{TC3} message was transmitted, the CAM waits at the DCC queue ($t_{q1}$) until it is dequeued and transmitted ($t_{tx1}$). The message then reaches a remote ITS station, where it is decoded by the CA service. The latency of this message ($t_{e2e1}$) is mainly dominated by the time it waited in the queues, and when the next CAM is received at $t_{tx2}$, this delay will add to the inter-packet gap to constitute the information age of the first message ($t_{age1}$).

Considering a system in steady state with two types of traffic with different \textcolor{black}{traffic classes and priorities (e.g, CAM with TC2 and another type of messages with TC3)}, where CAM generations occur at a regular rate (i.e., $t_{cam}$ is practically constant), the total message rate for a vehicle is modeled in equation \ref{equation:rates}:

\begin{equation}
\label{equation:rates}
\begin{split}
r_{total} = \frac{1}{t_{dcc}}; r_{cam} = \frac{1}{t_{cam}}; r_{tc3} = r_{total} - r_{cam}
\end{split}
\end{equation}

\textcolor{black}{where message rates ($r$) are defined for CAM ($r_{cam}$) and lower priority traffic ($r_{tc3}$), that add up to the total message rate for the vehicle ($r_{total}$). If  $t_{cam}$ is equal to $t_{dcc}$, $r_{tc3}$ approaches zero and \mbox{$r_{total} = r_{cam}$}. However, if $t_{cam}$ is larger than $t_{dcc}$, TC3 finds gaps to be transmitted, becoming a source for desynchronization between the interval provided by DCC to the CA service and the actual time a CAM will be dequeued. Both cases are exemplified in Fig. \ref{fig:dia1}}

The DCC mechanism at the Access layer dequeues a message according to the rate defined by channel occupation. After a message is transmitted \textcolor{black}{(i.e., at $t_{tx}$ in Fig. \ref{fig:dia1})}, the time at which the next message is dequeued (i.e., $t_{go}$) is updated to $t_{tx}$ plus a time between 25 and 1000 milliseconds (i.e., $t_{dcc}$), as indicated in~\cite{etsiNewDcc} Annex B. The transmission of a lower priority message can potentially delay higher priority traffic. An example can be illustrated when a CAM generation occurs instants after a \textcolor{black}{TC3} message was transmitted, which will cause the CAM to wait for the next gate opening at the DCC queues. 

Another source of delays at the DCC queues can emerge when $t_{cam}=t_{dcc}$ and the feedback from DCC at the CA service is not synchronized with $t_{go}$ (lower part of \mbox{Fig. \ref{fig:dia1}}), and thus a generated CAM must wait until DCC allows its transmission. This waiting time at the queue ($t_{q}$) is carried over to the end-to-end delay metric ($t_{e2e}$). This effect, even when not considering losses in the channel, adds to the age of the information a vehicle has about their neighbors, leading to undesirable ramifications such as tracking errors. 

The information provided to the CA service by DCC through the Management Entity \cite{etsiCAM} consists only of the allowed message rate, but it does not specify the instant when the next transmission will occur, and thus a CAM can be generated at any time between two consecutive transmissions. Due to this desynchronization, $t_{q}$ can be of any length from zero to $t_{dcc}$, and the relation between this waiting time at the queue and the time interval provided by DCC is a uniform distribution as expressed in equation \ref{equation:e2e}:

\begin{equation}
\label{equation:e2e}
\begin{split}
T_q \sim U(0,t_{dcc}) \therefore \overline{t_{q}} = \frac{t_{dcc}}{2} \\
\end{split}
\end{equation}

This becomes an issue when $t_{dcc}$ is high. Besides affecting the end-to-end delay ($t_{e2e}$), the waiting time at the queue affects another related metric: the information age ($t_{age}$) of CAM messages at remote stations. Information age is the time difference between the generation time-stamp in the most recent CAM received from a neighbor and the time a new message is received. In a channel with no losses, the minimum age of the information of the last CAM at the moment of the reception of a new one is expressed in equation \ref{equation:age}:
\begin{equation}
\label{equation:age}
t_{age(n-1)} = t_{q(n)} + t_{cam(n)}
\end{equation}

where $t_{q(n)}$ is the waiting time for the CAM being received, and $t_{cam(n)}$ is the time elapsed between two CAM generations (i.e., the intended time between updates). Longer times waiting at the queues have an effect on information age,  since time offsets can be as high as $2 \cdot t_{dcc}$.

\textcolor{black}{Equation~\ref{equation:age} does not consider some additional contributions to end-to-end delay: 1) waiting time in the EDCA queue, regulated by the IEEE 802.11 medium access control mechanism; 2) the packet transmission time; and 3) propagation delay. The sum of these delays has a value close to 1\,ms (see ~\cite{access}) and is, therefore, negligible compared with the delays we are analyzing in this letter.}

\subsection{Generate-on-Time (GoT)}
We propose the use of another parameter in combination with the ones defined in the rules established by \cite{etsiCAM}. \textcolor{black}{Using the existing architecture~\cite{etsiCross}, the interface to the Management Entity will provide, besides the message rate $t_{dcc}$, the time the next transmission will occur ($t_{go}$). Then, when conditions 1 or 2 would trigger a CAM generation in the CA service, another condition is checked: if $((t_{go} - t) - \varepsilon) \le 0$, where $t$ is the current time and $\varepsilon$ is greater than the time required to create a CAM message. If this condition is not met, the CAM is not generated at this instant, but the current time is stored ($t$), along with the current values for position, acceleration, and heading (i.e., $D$, for dynamics). The CA service will wait for the time provided by the Management Entity (e.g., sleeping for $(t_{go} - t) - \varepsilon$ ). After this wait, when $t_{go}$ allows for the CAM to be generated at the instant $t'$, it will include the most recent values for dynamics ($D'$)}. However, the next CAM generation will be referred to the stored values (i.e., $t_{cam}$ will be calculated using $t$ instead of $t'$, and shifts in dynamics will be compared to $D$ instead of $D'$), keeping average $t_{cam}$ within the values it would have had with the standardized CAM generation rules. CAM messages using GoT, time-stamped with $t'$ and including $D'$, will update neighbors using the most recent information from the vehicle data provider. 

\textcolor{black}{The advantages of delaying CAM generation depend on the ability of obtaining updated location information in the vehicle. Global Navigation Satellite Systems (GNSS) have a minimum time between measurements. However, because the CAM generation is not synchronized with the GNSS measurements, even small delays can allow obtaining an updated location position. Moreover, the use of additional sensors (inertial sensors) for tracking functionality enables a continuous reading of vehicle position between GNSS measurements.}

\begin{figure}[t]
	\centering
	\includegraphics[width=200px,clip,keepaspectratio]{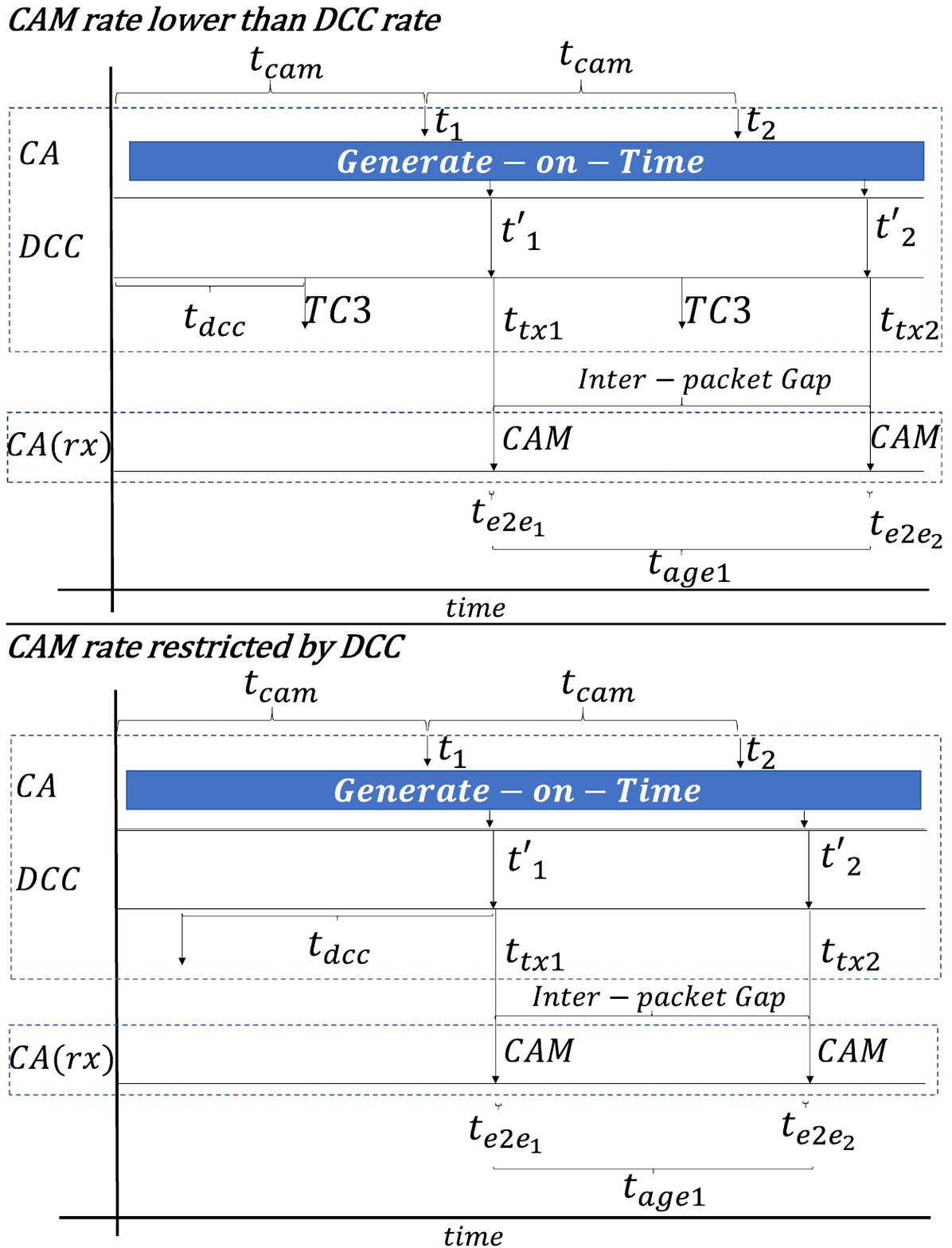}
	\caption{CAM generation with Generate-on-Time}
	\label{fig:dia2}
\end{figure}

Fig. \ref{fig:dia2} models the behavior of GoT, which acts as a layer that registers when a CAM needs to be generated, but completes the generation \textcolor{black}{once $t_{go}$ approaches}, using a time buffer ($\varepsilon$) to guarantee the message will be ready to be dequeued by the DCC mechanism. A new $t_{cam}$ will be calculated taking $t_n$ as a reference. When comparing Fig. \ref{fig:dia1} and Fig. \ref{fig:dia2}, it is shown that CAM messages triggered at $t_n$ will not be dequeued until $t_{tx}$ for both approaches. However, GoT delays the fulfillment of this trigger in order to compensate for $t_{q}$, improving $t_{e2e}$ and $t_{age}$. Regardless of the mechanism being used, transmissions will occur at the same time ($t_{tx}$), and neighbors will receive CAM messages not only at the same rate, but also at the same time (i.e., Inter-packet gaps are equal for both approaches), but GoT guarantees that the information in CAM messages is the most up to date possible.

\textcolor{black}{TC3 traffic} is not the only source of desynchronization, for example, for cases where there is traffic with higher priority than CAMs (e.g., high-priority DENM messages), CAMs triggered with GoT or ETSI rules are equally affected: a TC0 or TC1 message will update $t_{go}$ and delay the dequeuing of a CAM or, in the case of GoT, it can possibly delay its generation. For both cases, the next CAM will be sent the next time $t_{go}$ allows it. However, GoT synchronizes the next CAM generation with the DCC gate opening. This is an important property of GoT, since in ETSI CAM, any event that creates desynchronization will not be corrected until $t_{cam}>t_{dcc}$ and no other type of traffic is transmitted.

\section{Simulation Results}
\label{sec:simulation}
In order to evaluate the performance of ETSI CAM and GoT, we performed two types of experiments: 1) \textcolor{black}{a static scenario} where CAM generation is triggered at fixed rates (0.1 and 0.3 s), but restricted by feedback from DCC at the Facilities layer; and 2) a highway with five vehicle densities ranging from 10 to 50 vehicles/km per lane. Simulations are performed on Artery \cite{Artery}, with the parameters shown in Table~\ref{tbl:simpars}. The value we used for $\varepsilon$ was 15\,ms, which is a conservative value that allows for the construction of a CAM and which can be adjusted accordingly in real implementations.

\begin{table}[tb]
	\centering
	\caption{Simulation parameters}
	\label{tbl:simpars}
	\begin{tabular}{| l | l |}
		\hline
		\textbf{Parameter}  & \textbf{Values} \\
		\hline
		Access layer protocol & ITS-G5 (IEEE 802.11p) \\
		Data rate & 6 Mbit/s \\
		Transmit power & 126 mW \\
		Channel bandwidth & 10 MHz at 5.9 GHz \\
		Path loss model & Two-Ray Interference Model \\
		Sensing range & 750 m \\
		DCC mechanism & ETSI Adaptive Approach DCC \\
		Packet Size (including certificates) & CAM: 335 Bytes, TC3: 332 Bytes \\
		GoT $\varepsilon$ & 15\,ms \\
		\hline
	\end{tabular}
\end{table}

\subsection{Static Scenario}
The static scenario \textcolor{black}{consists of 300 vehicles deployed equidistantly within 200} meters of each other, well into communication range. To evaluate cases of multi-traffic coexistence, one where \textcolor{black}{TC3} traffic is allowed to be transmitted and one where it is not, we generate traffic at two different rates: one case with CAM messages attempting to be generated every 100\,ms and another with 300\,ms between generations. In both cases, CAM generations are restricted by the feedback from DCC Facilities, e.g., a CAM generation would be triggered every 100\,ms, but it will be limited by the $t_{dcc}$ parameter.

\begin{figure}[t]
	\begin{subfigure}{.5\textwidth}
	\centering
	\includegraphics[width=200px,clip,keepaspectratio]{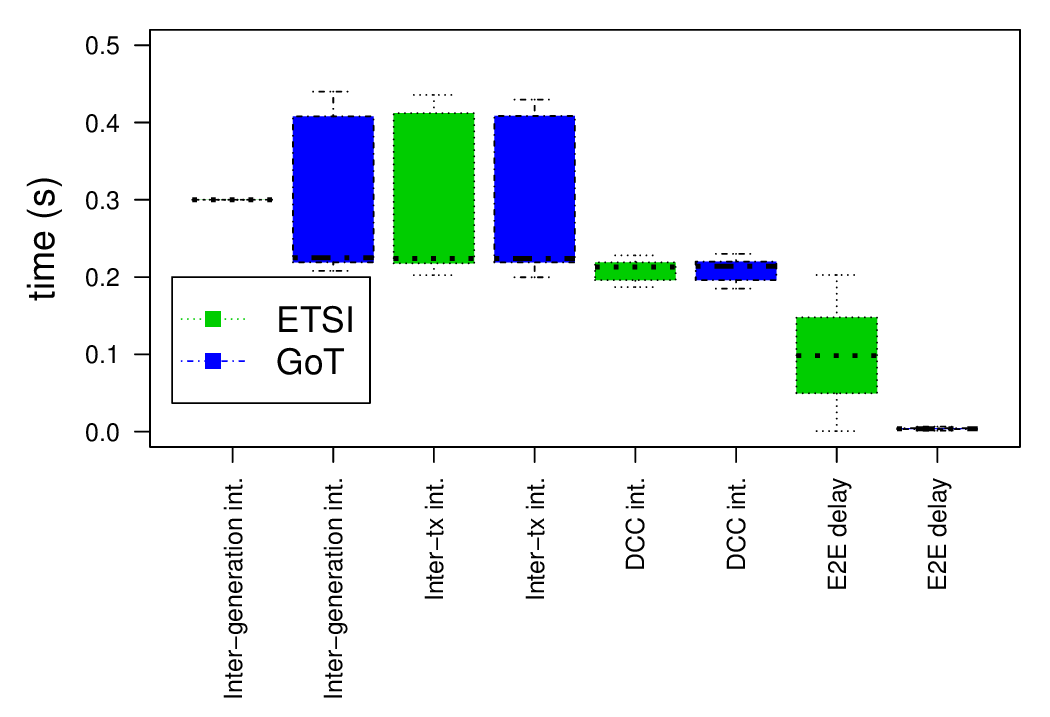}
	\caption{CAM generation triggered every 300\,ms}
	\label{fig:300_ms}
\end{subfigure}
	\begin{subfigure}{.5\textwidth}
	\centering
	\includegraphics[width=200px,clip,keepaspectratio]{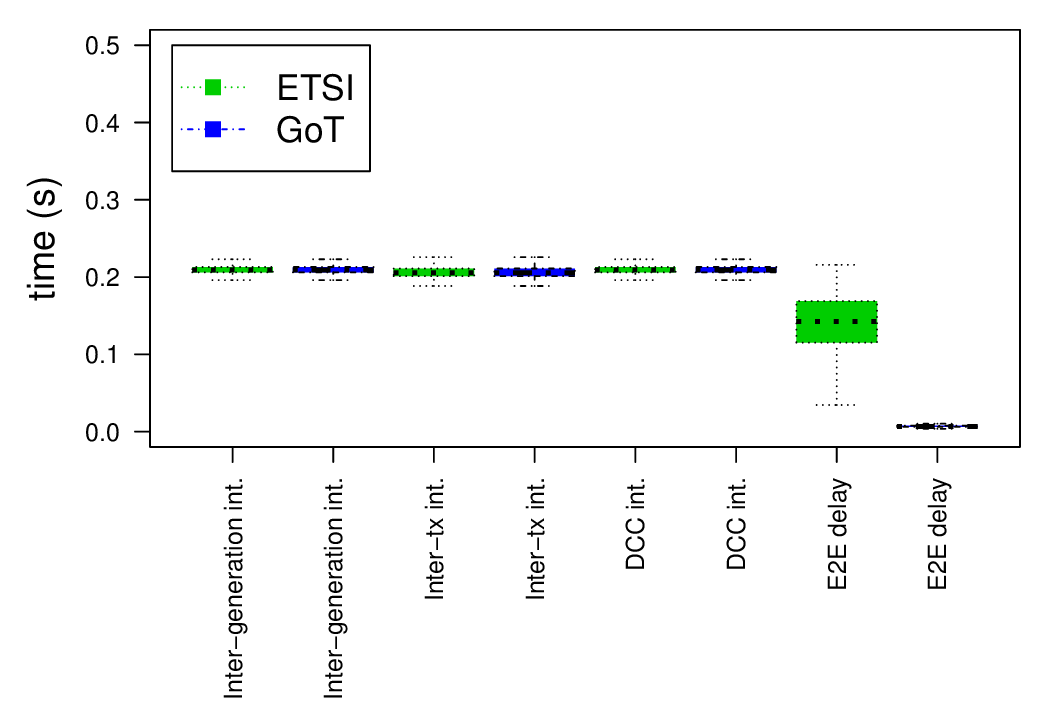}
	\caption{CAM generation triggered every 100\,ms}
	\label{fig:100_ms}
\end{subfigure}
	\caption{Comparison between ETSI CAM and Generate-on-Time in a static scenario}
\end{figure}

To evaluate the case where $t_{dcc} < t_{cam}$, CAM traffic was triggered every 300\,ms, well above the 200\,ms inter-generation interval allowed by DCC. Fig. \ref{fig:300_ms} shows that there is a greater variation between CAM generations with GoT, where the values for inter-generation intervals (i.e., the time between two CAM arrivals at the DCC queues) range from approximately 200 to 400\,ms, while ETSI CAM generations are at a fixed rate of 300\,ms. However, when CAM messages go down the stack, actual transmissions occur at the same time for both approaches: every 200 or 400\,ms for both algorithms, following a bimodal distribution. 

\textcolor{black}{Differences can be seen in the end-to-end delay, where values for the ETSI CAM algorithm follow a uniform distribution with an average value of $\frac{t_{dcc}}{2}$, while GoT keeps average delay to a minimum, close to $\varepsilon$, due to the mechanism's behavior.} This means that, even when messages are generated at a constant rate by the ETSI CAM algorithm, they are transmitted at different intervals, while GoT tries to synchronize generation and transmission instants, in order to avoid waiting times at the DCC queues and thus lowering \mbox{end-to-end} delay. To explore the case where $t_{dcc} = t_{cam}$ (i.e., when the $t_{dcc}$ parameter delays generations), CAM traffic was triggered every 100\,ms. However, it was not allowed to be transmitted until the interval given by DCC Facilities allowed the generation (on average, $t_{dcc} = 200\,ms$). Fig. \ref{fig:100_ms} once again shows that messages generated using GoT maintain low values for end-to-end delay.

\subsection{Road Scenarios}
The road scenarios are deployed on an a 7.75 km oval road with eight lanes — four in each direction — that has both long straight stretches and curves at the edges. Vehicle densities range from 10 to 50 vehicles/km per lane, and measurements are taken on the center of a straightaway. While the sensing range is around 750 m, only measurements from cars within 400 m of each other are considered. This distance was chosen because it is longer than the safety \textcolor{black}{distance for a driver to react to danger.} 

\begin{figure}[t]
	\centering
	\includegraphics[width=200px,clip,keepaspectratio]{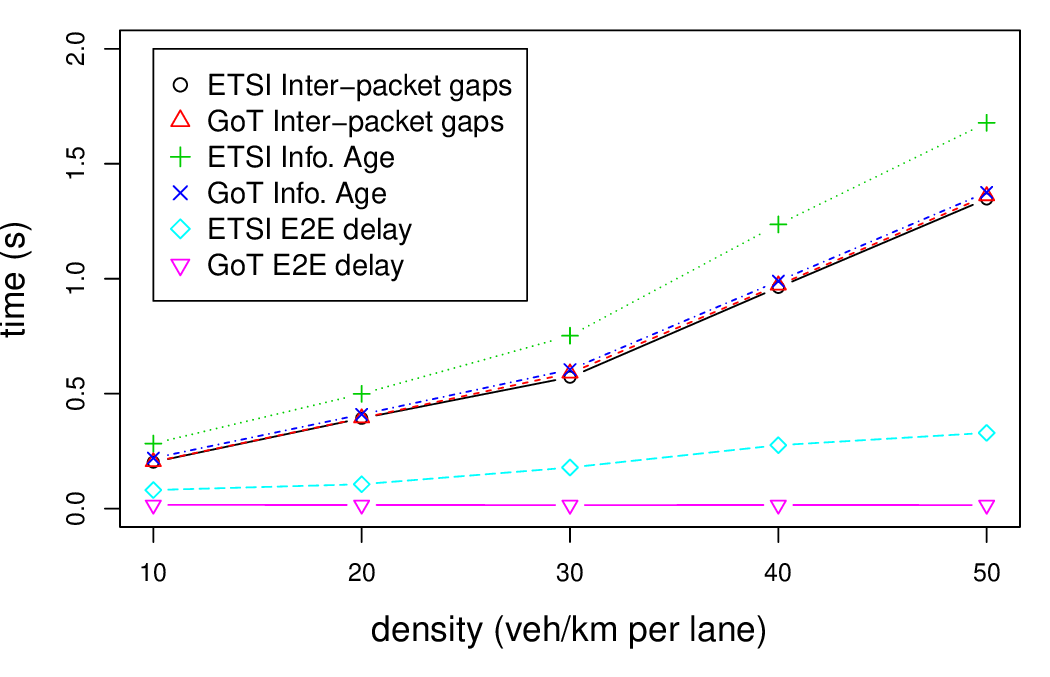}
	\caption{Comparison between ETSI CAM and Generate-on-Time in a road scenario ($d\leq400m$)}
	\label{fig:densities}
\end{figure}

Fig. \ref{fig:densities} shows that, even when GoT changes the CAM generation pattern, CAM messages are transmitted at the same rate and at the same time than CAM messages generated following the current ETSI rules. Average inter-packet gaps (IPG) are the same for both approaches, and they perform similarly on this metric at different densities, achieving the same performance due to the fact that transmissions happen at the same time, but ETSI CAM messages wait longer at the DCC queues. This waiting time is reflected on the Information Age metric, that shows the effect end-to-end delay has on the accuracy of the information a vehicle has about its neighbors. \textcolor{black}{Furthermore, even the smaller differences in end-to-end delay seen at the two lowest densities (63 and 89\,ms for densities of 10 and 20 veh/km per lane, respectively) give a better chance for the GNSS to provide a newer reading, even without position augmenting functions. Higher differences in \mbox{end-to-end} delay translate to errors that can affect safety, since the average end-to-end delay at the highest density is 302\,ms, which for this scenario (with an average speed of 14.27 m/s) translates to an error of 4.3 m, above the required position accuracy of many safety applications~\cite{etsiBSA}.}

\section{Conclusion}
\label{sec:conclusion}
We presented an analysis of the effect of multi-traffic on the end-to-end delay and information age metrics of CAM messages, by providing an analytical model and empirical results. In the simulations, we used CAM generation following the rules established by ETSI, and CAM messages coexisted with lower-priority traffic. Due to the desynchronization between the CA service and $t_{go}$, caused by the transmission of different types of messages, the DCC mechanism causes CAM traffic to wait in the queue after being generated and before being transmitted, adding to both the end-to-end delay of a message and the age of the information neighbors have about a station.

To deal with this effect, we proposed \textcolor{black}{Generate-on-Time} --- an addition to the CAM generation algorithm where the Management entity from the ETSI ITS-G5 protocol stack provides the time of the next dequeuing event (i.e., $t_{go}$), in order to generate CAMs when the DCC mechanism allows their transmission. We demonstrated that the use of GoT lowers end-to-end delay and information age to a minimum, while keeping the frequency at which CAM messages are transmitted at the same rate as the ETSI standardized CAM generation algorithm.

\bibliographystyle{IEEEtran}
\bibliography{references}

\begin{thebibliography}{10}
\providecommand{\url}[1]{#1}
\csname url@samestyle\endcsname
\providecommand{\newblock}{\relax}
\providecommand{\bibinfo}[2]{#2}
\providecommand{\BIBentrySTDinterwordspacing}{\spaceskip=0pt\relax}
\providecommand{\BIBentryALTinterwordstretchfactor}{4}
\providecommand{\BIBentryALTinterwordspacing}{\spaceskip=\fontdimen2\font plus
\BIBentryALTinterwordstretchfactor\fontdimen3\font minus
  \fontdimen4\font\relax}
\providecommand{\BIBforeignlanguage}[2]{{%
\expandafter\ifx\csname l@#1\endcsname\relax
\typeout{** WARNING: IEEEtran.bst: No hyphenation pattern has been}%
\typeout{** loaded for the language `#1'. Using the pattern for}%
\typeout{** the default language instead.}%
\else
\language=\csname l@#1\endcsname
\fi
#2}}
\providecommand{\BIBdecl}{\relax}
\BIBdecl

\bibitem{etsiCAM}
{European Telecommunications Standards Institute (ETSI)}, ``{Intelligent
  Transport Systems (ITS); Vehicular communications; Basic set of applications;
  Part 2: Specification of Cooperative Awareness basic service},'' {} {ETSI EN
  302 637-2 - V1.4.1}, April 2019.

\bibitem{CAMboban}
M.~{Boban} and P.~M. {d'Orey}, ``{Measurement-based evaluation of cooperative
  awareness for V2V and V2I communication},'' in \emph{2014 IEEE Vehicular
  Networking Conference (VNC)}, 2014, pp. 1--8.

\bibitem{CAMerlangen}
D.~{Eckhoff}, N.~{Sofra}, and R.~{German}, ``{A performance study of
  cooperative awareness in ETSI ITS G5 and IEEE WAVE},'' in \emph{{2013 10th
  Annual Conference on Wireless On-demand Network Systems and Services
  (WONS)}}, 2013, pp. 196--200.

\bibitem{CAMe2e}
M.~A. {Javed} and E.~{Ben Hamida}, ``{Measuring safety awareness in cooperative
  ITS applications},'' in \emph{{2016 IEEE Wireless Communications and
  Networking Conference}}, 2016, pp. 1--7.

\bibitem{etsiNewDcc}
{European Telecommunications Standards Institute (ETSI)}, ``{Intelligent
  Transport Systems (ITS); Decentralized congestion control mechanisms for
  Intelligent Transport Systems operating in the 5 GHz range; Access layer
  part},'' {} {ETSI TS 102 687 - V1.2.1}, April 2018.

\bibitem{etsiCross}
------, ``{European Telecommunications Standards Institute (ETSI),
  “Intelligent Transport System (ITS); Cross Layer DCC Management Entity for
  operation in the ITS G5A and ITS G5B medium},'' {} {ETSI TS 103 175 -
  V1.1.1}, June 2015.

\bibitem{Sepulcre2016}
M.~Sepulcre, J.~Gozalvez, O.~Altintas, and H.~Kremo, ``Integration of
  congestion and awareness control in vehicular networks,'' \emph{Ad Hoc
  Networks}, vol.~37, 2 2016.

\bibitem{dccHarri}
I.~{Khan} and J.~{H\"arri}, ``{Integration Challenges of Facilities-Layer DCC
  for Heterogeneous V2X Services},'' in \emph{2018 IEEE Intelligent Vehicles
  Symposium (IV)}, 2018, pp. 1131--1136.

\bibitem{Harri2018}
------, ``Flexible packet generation control for multi-application v2v
  communication,'' in \emph{2018 IEEE 88th Vehicular Technology Conference
  (VTC-Fall)}, 2018, pp. 1--5.

\bibitem{gunther2018}
H.-J. Günther, R.~Riebl, L.~Wolf, and C.~Facchi, ``The effect of decentralized
  congestion control on collective perception in dense traffic scenarios,''
  \emph{Comput. Commun.}, vol. 122, pp. 76 -- 83, 2018.

\bibitem{Letter}
I.~{Soto}, O.~{Amador}, M.~Urue{\~{n}}a, and M.~Calderon, ``{Strengths and
  Weaknesses of the ETSI Adaptive DCC Algorithm: A Proposal for Improvement},''
  \emph{IEEE Communications Letters}, vol.~23, no.~5, pp. 802--805, 2019.

\bibitem{access}
O.~Amador, I.~Soto, M.~Calderon, and M.~Urue{\~{n}}a, ``{Experimental
  Evaluation of the {ETSI} {DCC} Adaptive Approach and Related Algorithms},''
  \emph{{IEEE} Access}, vol.~8, pp. 49\,798--49\,811, 2020.

\bibitem{etsiDP}
{European Telecommunications Standards Institute (ETSI)}, ``{Intelligent
  Transport Systems (ITS); Harmonized Channel Specifications for Intelligent
  Transport Systems operating in the 5 GHz frequency band},'' {} {ETSI TS 102
  724 - V1.1.1}, October 2012.

\bibitem{etsiGeoNetworking}
------, ``{Intelligent Transport Systems (ITS); Vehicular communications;
  GeoNetworking; Part 4: Geographical addressing and forwarding for
  point-to-point and point-to-multipoint communications; Sub-part 2:
  Media-dependent functionalities for ITS-G5},'' {} {ETSI TS 102 636-4-2 -
  V1.2.1}, April 2020.

\bibitem{Artery}
R.~{Riebl}, H.~{G\"unther}, C.~{Facchi}, and L.~{Wolf}, ``{Artery: Extending
  Veins for VANET applications},'' in \emph{{2015 International Conference on
  Models and Technologies for Intelligent Transportation Systems (MT-ITS)}},
  2015, pp. 450--456.

\bibitem{etsiBSA}
{European Telecommunications Standards Institute (ETSI)}, ``{Technical Report.
  Intelligent Transport Systems (ITS);. Vehicular Communications;. Basic Set of
  Applications;. Definitions},'' {} {TR 102 638 - V 1.1.1}, June 2009.

\end{thebibliography}

\ifCLASSOPTIONcaptionsoff
  \newpage
\fi

\end{document}